\documentclass[aps,prl,single-column,groupaddress,superscriptaddress,amsmath,amssymb,graphicx]{revtex4-2}
\usepackage[utf8]{inputenc}
\usepackage{amsmath,amssymb,bm,amsfonts}
\usepackage{graphicx}
\usepackage{subfigure}
\usepackage{cancel}
\usepackage{booktabs}
\usepackage{bbold}
\usepackage{booktabs}
\usepackage{color}
\usepackage{ulem}
\usepackage{mathtools}
\usepackage{xcolor}
\usepackage{dsfont}
\usepackage{xparse}

\newcommand*\colvec[3][]{
    \begin{pmatrix}\ifx\relax#1\relax\else#1\\\fi#2\\#3\end{pmatrix}
}

\ExplSyntaxOn
\NewDocumentCommand{\mref}{m}{\quinn_mref:n {#1}}
\seq_new:N \l_quinn_mref_seq
\cs_new:Npn \quinn_mref:n #1
 {
  \seq_set_split:Nnn \l_quinn_mref_seq { , } { #1 }
  \seq_pop_right:NN \l_quinn_mref_seq \l_tmpa_tl
  ( 
  \seq_map_inline:Nn \l_quinn_mref_seq
    { \ref{##1},\nobreakspace } 
  \exp_args:NV \ref \l_tmpa_tl 
  ) 
 }
\ExplSyntaxOff

\date{\today}

\begin{document}
\title{Effect of delay on the emergent stability patterns in Generalized Lotka-Volterra ecological dynamics}
\author{Meghdad Saeedian}
\affiliation{Dipartimento di Fisica ``G. Galilei'', Università di Padova, Via Marzolo 8, 35131 Padova, Italy}
\author{Emanuele Pigani}
\affiliation{Dipartimento di Fisica ``G. Galilei'', Università di Padova, Via Marzolo 8, 35131 Padova, Italy}
\affiliation{Stazione Zoologica Anton Dohrn, Villa Comunale, 80121 Naples, Italy. }
\author{Amos Maritan}
\affiliation{Dipartimento di Fisica ``G. Galilei'', Università di Padova, Via Marzolo 8, 35131 Padova, Italy}
\author{Samir Suweis}
\affiliation{Dipartimento di Fisica ``G. Galilei'', Università di Padova, Via Marzolo 8, 35131 Padova, Italy}
\affiliation{Padova Neuroscience Center, University of Padova, Padova, Italy}
\author{Sandro Azaele}
\affiliation{Dipartimento di Fisica ``G. Galilei'', Università di Padova, Via Marzolo 8, 35131 Padova, Italy}

\begin{abstract}
\noindent
Understanding the conditions of feasibility and stability in ecological systems is a major challenge in theoretical ecology. The seminal work of May in 1972 and recent developments based on the theory of random matrices have shown the existence of emergent universal patterns of both stability and feasibility in ecological dynamics. However, only a few studies have investigated the role of delay coupled with population dynamics in the emergence of feasible and stable states. In this work, we study the effects of delay on Generalized Loka-Volterra population dynamics of several interacting species in closed ecological environments. First, we investigate the relation between feasibility and stability of the modeled ecological community in the absence of delay and find a simple analytical relation when intra-species interactions are dominant. We then show how, by increasing the time delay, there is a transition in the stability phases of the population dynamics: from an equilibrium state to a stable non-point attractor phase. We calculate analytically the critical delay of that transition and show that it is in excellent agreement with numerical simulations. Finally, we introduce a measure of stability that holds for out of equilibrium dynamics  and we show that in the oscillatory regime induced by the delay stability increases for increasing ecosystem diversity.
\end{abstract}

\maketitle

\section{introduction}

A fascinating aspect in the study of biological and ecological systems is the emergence of ubiquitous patterns that do not depend on the specific details of the system under study \cite{may1972will,may2019stability,volkov2007patterns,suweis2013emergence,azaele2016statistical,peruzzo2020spatial,gupta2021effective}
In this context, one of the main problems of theoretical ecology is the search for key mechanisms leading to the emergence and maintenance of biodiversity. The conditions for many species to coexist in spite of changing environment or perturbations are tightly connected to the problem of understanding when and how ecological systems are feasible (i.e., all solutions are positive at large times) and asymptotically stable (i.e., the real parts of all the eigenvalues of the Jacobian are negative).

The stability of ecosystems is mainly determined by some driving features, including diversity (number of species), the type of ecological interactions (antagonistic, competitive or mutualistic) among species, their strength and network structure, and sensitivity to environmental perturbation. Diversity is probably one of the easiest components that can be measured empirically \cite{loreau2001biodiversity,mccann2000diversity} and, along with the density of the species interactions in the community (known as connectance), it has been considered a standard indicator of ecosystem complexity. 

Before the 1970, ecologists such as Elton \cite{elton2020ecology}, Odum \cite{odum1953fundamentals}, MacArthur \cite{macarthur1955fluctuations}, and many others believed that diversity enhances the stability of ecosystems. However, later theoretical studies suggested exactly the opposite, and more works often confirmed a disagreement between empirical and theoretical studies on the relation between diversity and stability \cite{ives2007stability}. This is known as stability–complexity debate and it has been initiated by the seminal work of May in 1972 \cite{may1972will}. In that work, May investigated the linear stability of a null ecological ecosystem with random interactions, and found an analytical result based on random matrix theory, indicating that the more complex the ecosystem is, the less stable it is. Many other works, including recent developments on generalizations of May's work \cite{allesina2012stability,suweis2014disentangling,grilli2016modularity} confirmed the original result of May.  Since then, the complexity-stability paradox has been tackled through two main approaches: some works argued that the stationary condition of ecological systems cannot be described by equilibrium points \cite{pimm1984complexity,mccann2000diversity}, hence suggesting a change of perspective on stability. This led to replacing asymptotic stability measures with alternative variables of interest (e.g. the coefficient of variation of the ecosystem population abundance). Several other studies, instead, focused on investigating the role of ecological function \cite{jansen2003complexity} and structure of food webs \cite{landi2018complexity,gravel2016stability}. 
A few studies analyzed the role of non-linearity of the ecological dynamics on ecosystem stability \cite{gibbs2018effect,fedeli2021nonlinearity} or on the possible effect of delay \cite{takeuchi1996global}. To our knowledge, the impact of non-linearity and delay on the stability of ecological systems remain elusive.

To tackle this problem, first we consider a classic interacting ecological model called Generalized Lotka-Volterra (GLV) system. The term "Generalized" refers to a model that containes an arbitrary mixture of ecological interactions, such as prey-predator (PP), mutualism (MU), competition (CO) or other types of interactions. Following previous work, we model the interactions network through a random matrix (RM) approach and assume no-pattern structure in the ecosystem, as in May seminal work.

We will then add a temporal delay into the dynamics of species' populations. The resulting delayed GLV equations with a few number of species (low diversity) have analytically been studied mostly for the case of Prey-Predator systems \cite{may1973time,song2005local,yan2006hopf,faria2001stability,takeuchi1996global}. These studies have investigated various implementations of the delay in the temporal dynamics of the prey-predator. Almost all variants indicate the emergence of a Hopf bifurcation from an equilibrium state to periodic solutions. Also, similar works have investigated other characteristics of the delayed prey-predator systems such as boundedness of solutions, persistence, local and global stabilities of equilibria, and the existence of nonconstant periodic
solutions \cite{yan2008hopf,beretta1996convergence,freedman1986stability,freedman1995uniform,he1996stability,ma1998stability,zhao1997global}.


Delayed interactions are intrinsic to a variety of systems and are realistic and ubiquitous features which ought to be incorporated in many population-dynamics models. Consider for example, a closed ecological environment like a small lake which contains phytoplankton (P), zooplankton (Z), fish (F), and some inorganic nutrients (N) as a limited resource \cite{takeuchi1996global}. The food chain is then: the phytoplankton consumes inorganic nutrients and is consumed by zooplankton. The fishes consume zooplankton. After these organisms (P,Z,F) die, the decomposers recycle the dead organic carcass to inorganic nutrients (N) after a certain delay time $\tau$. Therefore, the increase of N(t) depends not merely on the population P(t) at time t, but also on the population of individuals (P,Z,F) that died in the past, i.e., P(t-$\tau$), Z(t-$\tau$), and F(t-$\tau$). There have been developed a lot of mechanisms that may induce delay interactions in the ecosystems such as: maturation period (see Figure 3 of \cite{nicholson1954outline}), a gestation period \cite{scudo1971vito}, feeding times and hunger coefficients in predator-prey interactions \cite{caswell1972simulation}, replenishment or regeneration time for resources (e.g. of vegetation for grazing herbivores \cite{may1973time,may1974time}). \textit{One can easily imagine other causes of delays in population dynamics on various time scales: those caused by food storage of predators or gatherers, reaction times, threshold levels, etc...} \cite{cushing1977integro}. Finally, the presence of time delay may also depend on the spatial scale of observation: mean field equations (i.e., without explicit space) of spatially-extended systems may include distributed time delays, depending on the spatial scale that is implicit.  

The main objective of this work is to characterize the emergence of instability in delayed GLV ecosystems and in connection with the  complexity-stability debate. We first start by turning off the delay and characterize analytically the stability in non-delayed GLV. In this case we have found an analytical connection between feasibility, stability and diversity in the coexistence state, where the system is feasible and stable. When we turn on the delay, we find that it has a detrimental effect on the stability of the ecological dynamics. To be more precise, by gradually increasing the delay, we first observe a decrease in stability (or resilience) of the system (measured by the absolute value of the leading eigenvalue of the community matrix). At a certain critical delay the system experiences a Hopf bifurcation from an equilibrium state to an oscillatory one (with non-point attractors, including regular and irregular cyclic behaviors). In other words, for enough strong delay, we observe persistent oscillatory regimes that can not be predicted by the linear stability analysis.  We investigate such transition as a function of different delays and ecosystems complexity. We find that the critical delay of the bifurcation decreases by increasing the diversity, consistently with May results. By increasing the delay, the oscillatory regime persists until we observe numerical divergences in the trajectories of the populations. We suggest that this phenomenon is due to numerical instabilities. However, we do not prove the existence of bounded solutions for the delayed GLV equations. Finally, following \cite{pimm1984complexity,mccann2000diversity} we show that in the oscillatory regime induced by the delay, we can introduce a stability index for non-equilibrium systems measured as the coefficient of variability (CV) of the total population in the community. We find that in this case stability increases for increasing ecosystem diversity \cite{tilman1998diversity}. This result is in agreement with similar conclusions, albeit obtained with different mechanisms, which have been observed in theoretical and \cite{mccann2000diversity,huisman1999biodiversity}, experimental studies \cite{tilman1994biodiversity,tilman1996productivity,tilman1996biodiversity,van1998mycorrhizal,mcgrady1997biodiversity,mcgrady2000biodiversity,morin1995food,naeem1997biodiversity,naeem1998species,yachi1999biodiversity}. 

The paper is organized as follows. First, we present the theoretical frame work GLV in the absence and the presence of the delay. Next, we investigate the feasibility and stability of the GLV ecosystem in the absence of delay. Afterward, we start to suty the GLV in the presence of the delay. In the following we give some hints about the empirical perspective of stability in delayed GLV. The final section summarizes our main results.

\section{Theoretical Framework}

We consider a pool of $S$ interacting species with abundances $x=\{x_1,x_2,...,x_S\}$. Each species is characterized by an intrinsic growth rate, $r_i$, and by the interactions with other species that we here consider to have both an instantaneous and a delayed  effect of time $\tau$ on its abundance. Therefore, the equations governing the dynamics of $x_i$ read as
\begin{eqnarray}
\dot{x}_i(t)&=&x_i(t)\big(r_i+\sum_{j}a_{ij}x_j(t)+\sum_{j}b_{ij}x_j(t-\tau)\big),
 \label{CLV} 
\end{eqnarray}
which are known as delayed Generalized Lotka-Volterra (GLV) equations.
The matrix entries $a_{ij}$ and $b_{ij}$ measure the strength of the impact of species $i$ on species $j$, in the instantaneous and delayed interactions, respectively.

Real ecosystems with a large number of species include a mixture of all types of ecological interactions, such as predator-prey(PP), competitive (CO) and mutualistic (MU), and with a variable interactions strength. Typically such interactions are very difficult to infer from empirical data, and therefore, following the same spirit of the seminal work by May, we draw at random (quenched) both the intrinsic growth rate ($r_i$ from a Uniform distribution) and the non-diagonal components of the matrix $A$ and $B$ from a Normal distribution with probability $C \in [0;1]$, where $C$ is the so-called connectance, and 0 otherwise. The mean and standard deviation of the intrinsic growth rates are denoted by $\mu_r$ and $\sigma_r$, while the those for the strength of the interactions are denoted by ($\mu_A$, $\sigma_A$) and ($\mu_{B}$, $\sigma_{B}$). In this model, we consider the intra-species competition, represented by the diagonal component of the interaction matrices, to be constant, i.e.,  $a_{ii}=-d_A$ and $b_{ii}=-d_B$. The delay parameter, $\tau$, is one control parameter of the proposed model, and it is not random. 
In this work are interested to understand the effect of the delay on the feasible (i.e. non negative) and stable equilibria of the ecosystem as a function of the complexity ($S\cdot C$), given fixed all other ecological parameters (e.g. growth rates, interactions strengths).

Characterizing the stability and feasibility of a GLV system (Eqs.~\ref{CLV}) or similar ecological systems even in the absence of delay (i.e., $\tau=0$) is a non-trivial task. We note that When considering asymptotically equilibrium regimes, several type of stability can be defined \cite{ives2007stability}. Moreover, there has been a lot of effort to establish some sufficient analytical conditions to grant the stability of ecological community dynamics without delay \cite{case1979global,goh1977global,goh1978sector,takeuchi1980existence,takeuchi1978stability,takeuchi1978global,logofet2005stronger}. These sufficient conditions are mostly either established by Linear Stability Analysis (such as D-Stability), or proper Lyapunov functions (or other stronger conditions) \cite{logofet2005stronger}.

When we turn on the delay (i.e., $\tau>0$) things get complicated even further as the delay induces strong non-linearity on the dynamics given by Eqs.~\ref{CLV}. In this case we thus focus only on the local systems stability. 
Let $u$ be a small perturbation away from equilibrium state $x^\star$, then the linearized equation for $u(t)=x(t)-x^\star$ reads as
\begin{eqnarray}
\dot{u}_i &=& \sum_{j}\Tilde{A}_{ij} u_j(t)+\sum_{j}\Tilde{B} u_j(t-\tau),
 \label{LinDelay} 
\end{eqnarray} 
where $\tilde A$ and $\tilde B$ are the community matrices of the system due to the instantaneous and delayed interactions, respectively.  
Plugging the ansatz solution $u(t)=ve^{\lambda t}$ into Eq.~\ref{LinDelay}, where $\lambda$ is the eigenvalue of the system and $v$ is a constant vector in $\textbf{R}^S$, the $i$-th eigenvalue of the system, under the condition $[\tilde A$ , $\tilde B] =0$, satisfies the equation 
\begin{eqnarray}
\lambda_i&=&\tilde a_i+\tilde b_i e^{- \tau \lambda_i},
 \label{Lambda} 
\end{eqnarray} 
which can be equivalently expressed by means of the Lambert function \cite{corless1996lambertw} $W$ as
\begin{eqnarray}
\lambda_i&=&\tilde a_i+\frac{W(\tilde b_i \tau e^{-\tilde a_i \tau})}{\tau}
 \label{Lambda_Lambert} 
\end{eqnarray} 
where $\tilde a_i$ and $\tilde b_i$ are the eigenvalues of $\tilde A$ and $\tilde B$, respectively. Finally, notice that in general the eigenvalue here is a complex number, i.e., $\lambda_j=\xi_j+i\nu_j$ and $\tilde{b}_j=\alpha_j+i\beta_j$.

\section{Results}
\subsection{The feasibility and stability of the GLV ecosystem in the absence of delay}
As a starting point for the study of non-linear effects on large systems dynamics, let us begin by presenting some results on the feasibility and stability the classical GLV model without delay\cite{hofbauer1998evolutionary}. In particular, we present a criterion to quantify the probability for the steady state vector $x^\star$ to be feasible, i.e., to have all positive components, and we find that a correlation between feasibility and stability does exist, at least when the intra-species interactions are dominant with respect to the inter-species ones.

Let us consider Eq.\eqref{CLV} with $B=0$, that is, 
\begin{equation}
    \dot{x}_i(t)=x_i(t)\,g_i\left[x(t)\right] \equiv f_i\left[x(t)\right],
   \label{eq:LV1}
\end{equation}
where
\begin{equation}
   g_i\left[x(t)\right] = r_i + \sum_{j=1}^S a_{ij}x_j(t).
   \label{eq:LV2}
\end{equation}
Naturally, such a system admits up to $2^S$ different equilibria, since the equilibrium condition $f\left[x^\star\right]=0$ is satisfied by setting either $x_i^\star$ or $g_i\left[x^\star\right]$ to zero for each species. Thus, in principle one should study the feasibility and the stability of every equilibrium point, which becomes hard as the number of species growths. As we are interested in ecological models for the evolution of species' abundances or densities, we focus on the non-zero equilibria given by the condition $g\left[x^\star\right]=0$, that is, on $x^\star=-A^{-1}r$.
However, it is worth highlighting that the study of these equilibria  is meaningful only if they are feasible. Indeed, when some components of $x^\star$ are negative, the system will never reach such a state: starting with a positive initial condition $x(0)$, the abundance of each population $x_i$ can either increase or decrease, but it will never become negative, owing to the $x_i(t)$ factor in Eq.\eqref{eq:LV1}. Furthermore, we cannot claim anything about which other equilibrium the system will eventually reach, nor the final number of surviving species. In other words, if we compute $x^\star=-A^{-1}r$ and some components are negative, this implies neither that those represent species which will become extinct, nor that the number of surviving species corresponds to the number of positive components of $x^\star$, as explained through a simple example in Appendix \textbf{B}.

Hence, it is important to determine what conditions guarantee the feasibility of $x^\star$ and therefore we are allowed to interpret them as actual equilibrium populations. Although a general condition is lacking, Although recent works have investigated some conditions necessary for the feasible coexistence\cite{grilli2017feasibility}, a general condition is still lacking. However, herein we compute the feasibility probability when $A$ is a diagonally-dominant matrix, namely, when $d_A \gg \sigma_A$. Under this hypothesis, if we define $M$ as the off-diagonal part of $A$, i.e.,
$A = - d_A\, \mathds{I}+M$, then we can formally expand the inverse of the interaction matrix as 
\begin{equation}
A^{-1} = -\frac{1}{d_A} \sum_{k=0}^{\infty}\left(\frac{M}{d_A}\right)^k = -\frac{1}{d_A}  \left(\mathds{I} + \frac{M}{d_A}\right) + \text{less dominant terms}.
\end{equation}
Hence, at the first order, the steady states now read:
\begin{equation}
x^\star_i = \frac{r_i}{d_A} + \sum_{j=1}^S \frac{M_{ij}r_j}{d_A^2}
+ \text{less dominant terms}.
\label{eq:xstarapp}
\end{equation}
Since $r$ is a random variable and $M$ is a random matrix, $x^\star$ is a random vector and it is therefore possible to compute approximately the feasibility probability as a function of the system parameters. For instance, if the growth rates are constant and positive, $r_i=r>0$, and $M$ is a zero-diagonal, normal random matrix with zero mean, standard deviation $\sigma=\sigma_A$ and connectance $C$, then in the large size limit after some calculations (see Appendix \textbf{B}) we have
\begin{equation}
P_{\text{feas}}\left(d_A,\sigma_A,S,c\right)= \left\{\frac{1}{2}\text{erfc}\left[-\frac{d_A}{\sqrt{2C\left(S-1\right)}\sigma_A}\right]\right\}^S,
\label{eq:feas0}
\end{equation} where $\text{erfc}(\cdot)$ is the complementary error function \cite{temme2010error}. Remarkably, the feasibility probability decreases exponentially with the system size (at leading order in $S$, $P_{\text{feas}}\simeq 2^{-S}$).

\begin{figure}
    \centering
    \includegraphics[width=1\linewidth]{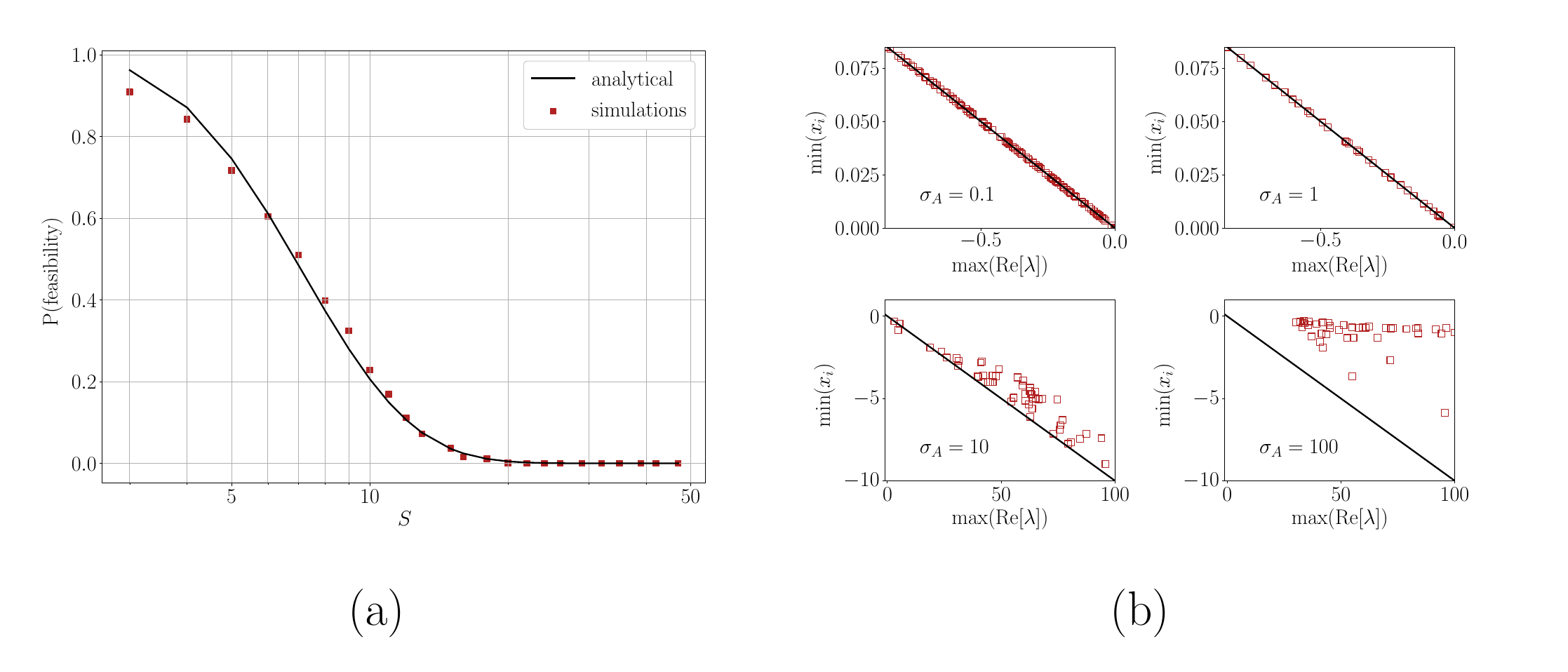}
    \caption{(a) Comparison between the theoretical prediction (solid line, Eq.\eqref{eq:feas0}) and numerical simulations (red squares) for the feasibility probability as a function of the system size $S$. Each red squared represents the fraction of feasible $x^\star$ obtained over 1000 realizations. We have considered $d_A=1$, $C=0.1$ and $\sigma_A=1$. (b) Numerical analysis of the relationship between the stability (given by the maximum real part eigenvalue max(Re[$\lambda$])) and the population of the rarest species (min($x^\star$)) for different values of $\sigma_A$. Red squares represent a mean over 100 realization, whereas the linear correlation (black line) is the first order of Eq.\eqref{eq:lambdaGLV1}. Such correlation holds only when $\sigma_A<d_A$ (upper panels). In each panel, $d_A=10$, $C=0.1$ and $S$ spans the interval from 10 to 200.}
    \label{fig:feas_stab}
\end{figure}

The second property that characterizes the equilibria is their (local) stability. This property can be studied by means of the spectrum of the Jacobian evaluated at a given fixed point $x^\star$, that is,
\begin{equation}
    J_{ij}(x^\star) \equiv \left.\frac{\partial f_i}{\partial x_j}\right|_{x^\star} = \delta_{ij}g_i(x^\star) + x_i^\star\left.\frac{\partial g_i}{\partial x_j}\right|_{x^\star} = \delta_{ij}g_i(x^\star) + x_i^\star a_{ij}.
\end{equation}
It is worth mentioning that the Jacobian relative to the trivial equilibrium $x^\star=0$ is $J_{ij}(0)=\delta_{ij} r_i$, and thus its stability depends only on the (sign) of the growth rates.
Instead, if we focus on the non-zero equilibrium $x^\star=-A^{-1}r$ and we assume that it is feasible, the Jacobian at $x^{\star}$ simply reads $J_{ij}=x_i^\star a_{ij}$. Although a general form for determing the eigenspectrum of $J$ is not known, it is possible to exploit perturbative expansion techniques for estimating the eigenvalues of the interaction matrix when $A$ is diagonally-dominant. Indeed, in this case a good approximation for the eigenvalues $\lambda_i$ reads
\begin{equation}
\lambda_i = -d_A x^\star_i  \left[1-\frac{1}{d_A}\sum_{j\neq i} A_{ij}A_{ji}\frac{x^\star_j}{x^\star_j - x^\star_i}  \right] 
+ \text{less dominant terms}.
\label{eq:lambdaGLV1}
\end{equation}
Thus, when $d_A\gg\sigma_A$ there is a simple relation at leading order between the eigenvalues and the steady states, and this case stability and feasibility are tightly link, as depicted in Fig.\ref{fig:feas_stab}. Of course, this is not true for general matrices that are not diagonally-dominant. Finally, notice that in the subsequent sections we stick to this condition.

\subsection{Delay as a bifurcation parameter: Hopf bifurcation from the asymptotically stable regime to the oscillatory regime}

In this section, we investigate the non-trivial solutions of the delayed GLV model with a large number of species.
To this purpose, we consider $A=0$ in order to study the pure effect of the delayed interaction. The system dynamics defined by Eq.\eqref{CLV} hence reduces to
\begin{equation}
    \dot{x}_i(t)=x_i(t)\big(r_i+\sum_{j}b_{ij}x_j(t-\tau)\big).
    \label{eq:dynamics}
\end{equation}
First, we calibrate the non-delayed system on the \textit{coexistence-of-species} state, where the solution of the system is feasible and stable. This state is easily obtained by considering $B \in \textbf{DiS}$ (Appendix \textbf{A}) and $r\in \textbf{R}^{S}_{+}$ \cite{servan2018coexistence}. Being in the coexistence region, we introduce the delay parameter and study the solutions of the system in the extended parameter space $(r,B,\tau)$. 

By definition, a Hopf bifurcation occurs when two complex conjugate eigenvalues of the community matrix, with non-zero imaginary part, simultaneously cross the imaginary axis into the right half-plane \cite{strogatz2018nonlinear}. Now, we investigate this condition in the community (Jacobian) matrix  of the linearized delayed GLV, and compare the results with the corresponding non-linear delayed GLV. 

\begin{figure}[ht]
\centering
\includegraphics[width=1\linewidth]{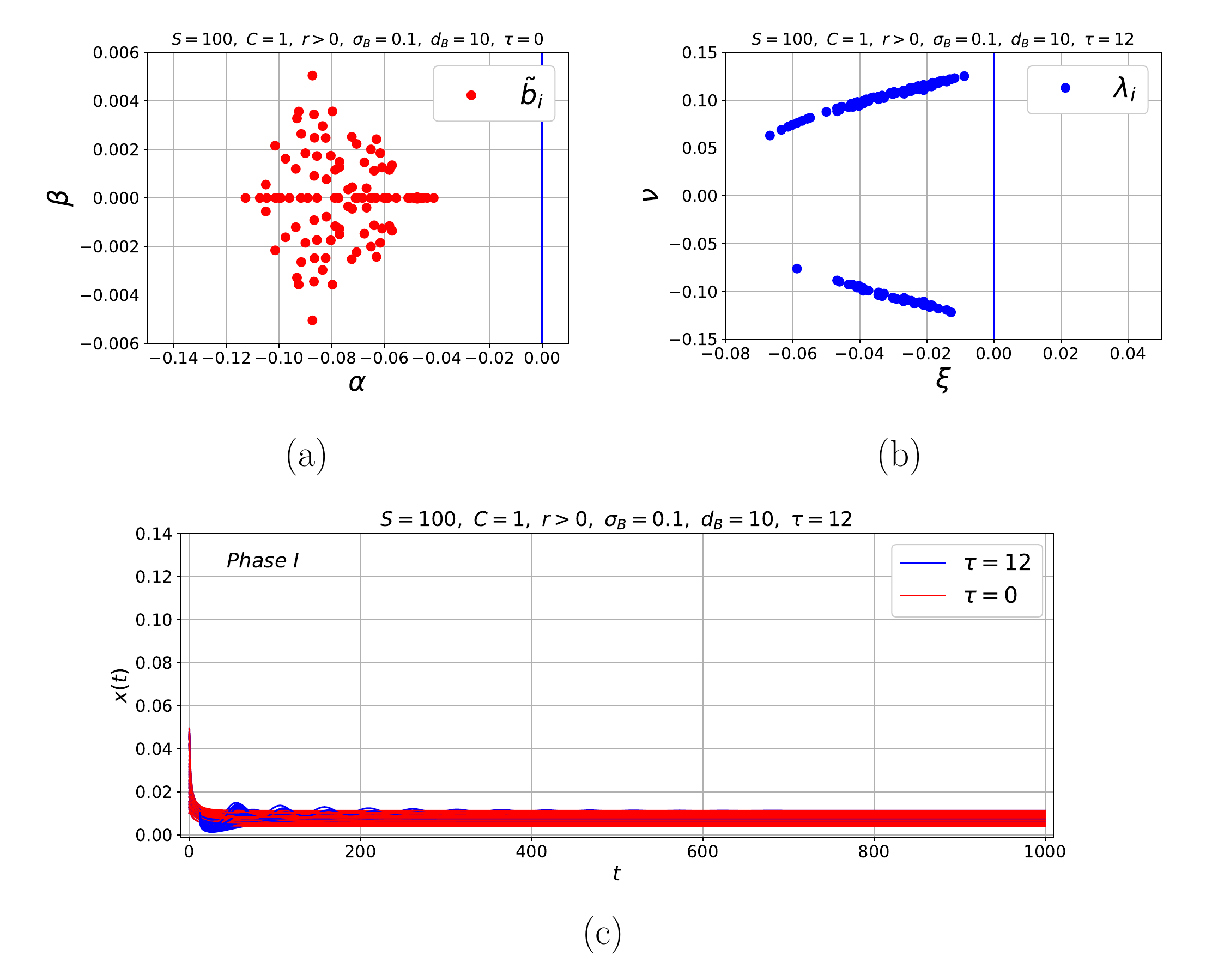}
\caption{Characterization of an asymptotically stable state for Eq.\eqref{eq:dynamics}. (a) Eigenspectrum of the community matrix $J=\text{diag}(-B^{-1}r)B$ for the non-delayed GLV system. (b) Eigenspectrum of the delayed GLV system for $\tau=12$ (see Eq.~\ref{Lambda_Lambert}). In both panels the real part of the leading eigenvalue is negative. Thus, as depicted by the results of the numerical integration of Eq.\eqref{eq:dynamics} in panel (c), the system is asymptotically stable both without (red line) and with (blue line) delay. Note, in all panels, $A=0$ and $B$ is a random matrix.
}
\label{nonlin-singleI}
\end{figure}
\begin{figure}[ht]
\centering
\includegraphics[width=1\linewidth]{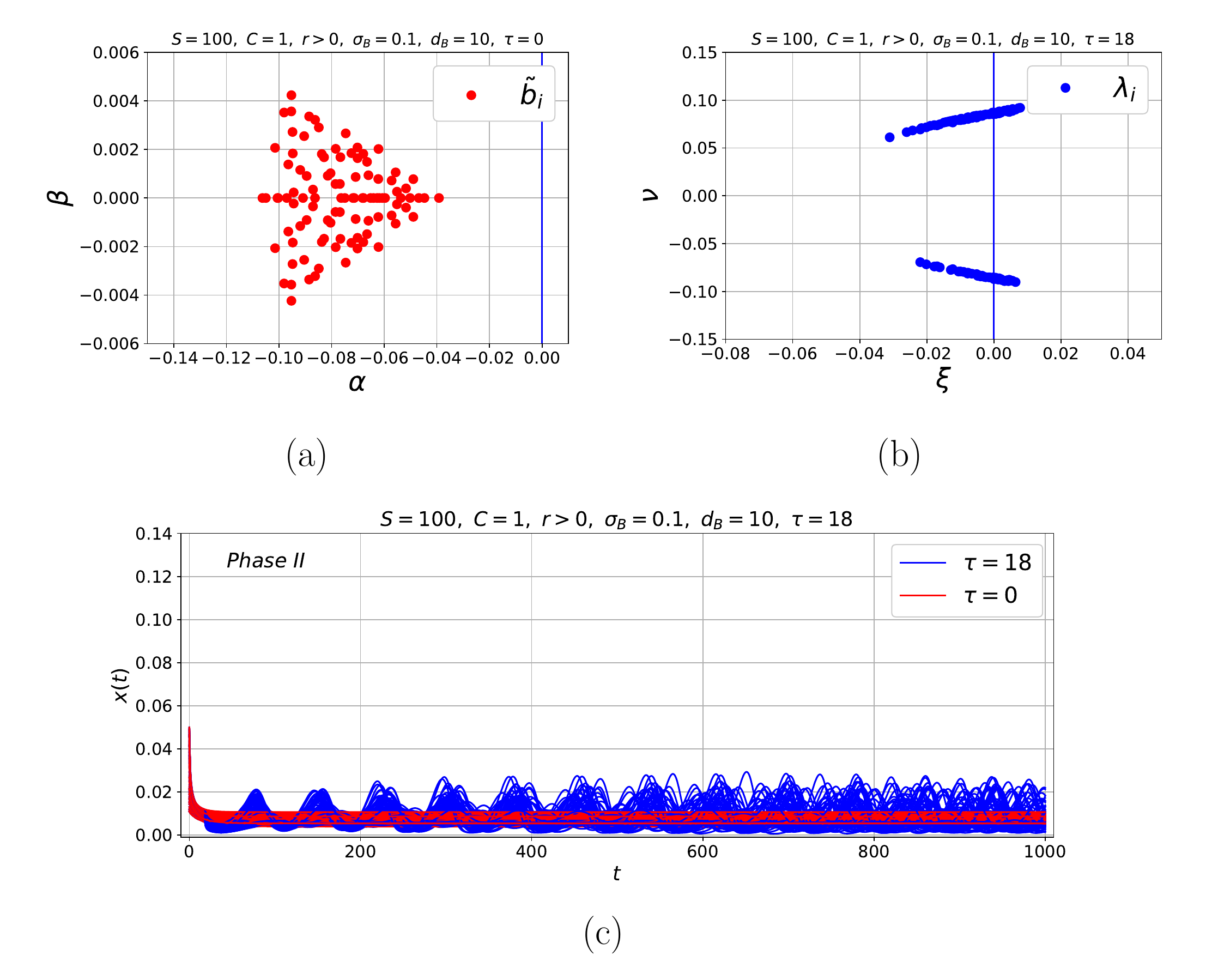}
\caption{Characterization of an oscillatory phase for Eq.\eqref{eq:dynamics}. (a) Eigenspectrum of the community matrix $J=\text{diag}(-B^{-1}r)B$ for the non-delayed GLV system. (b) Eigenspectrum of the delayed GLV system for $\tau=18$ (see Eq.~\ref{Lambda_Lambert}). While in panel (a) the real part of the leading eigenvalue is negative, it is instead positive in panel (b). Thus, as depicted by the results of the numerical integration of Eq.\eqref{eq:dynamics} in panel (c), the system without delay is asymptotically stable (red line), whereas it displays an oscillatory motion in the presence of delay (blue line). Note, in all panels, $A=0$ and $B$ is a random matrix.
}
\label{nonlin-singleII}
\end{figure}
\begin{figure}[ht]
\centering
\includegraphics[width=1\linewidth]{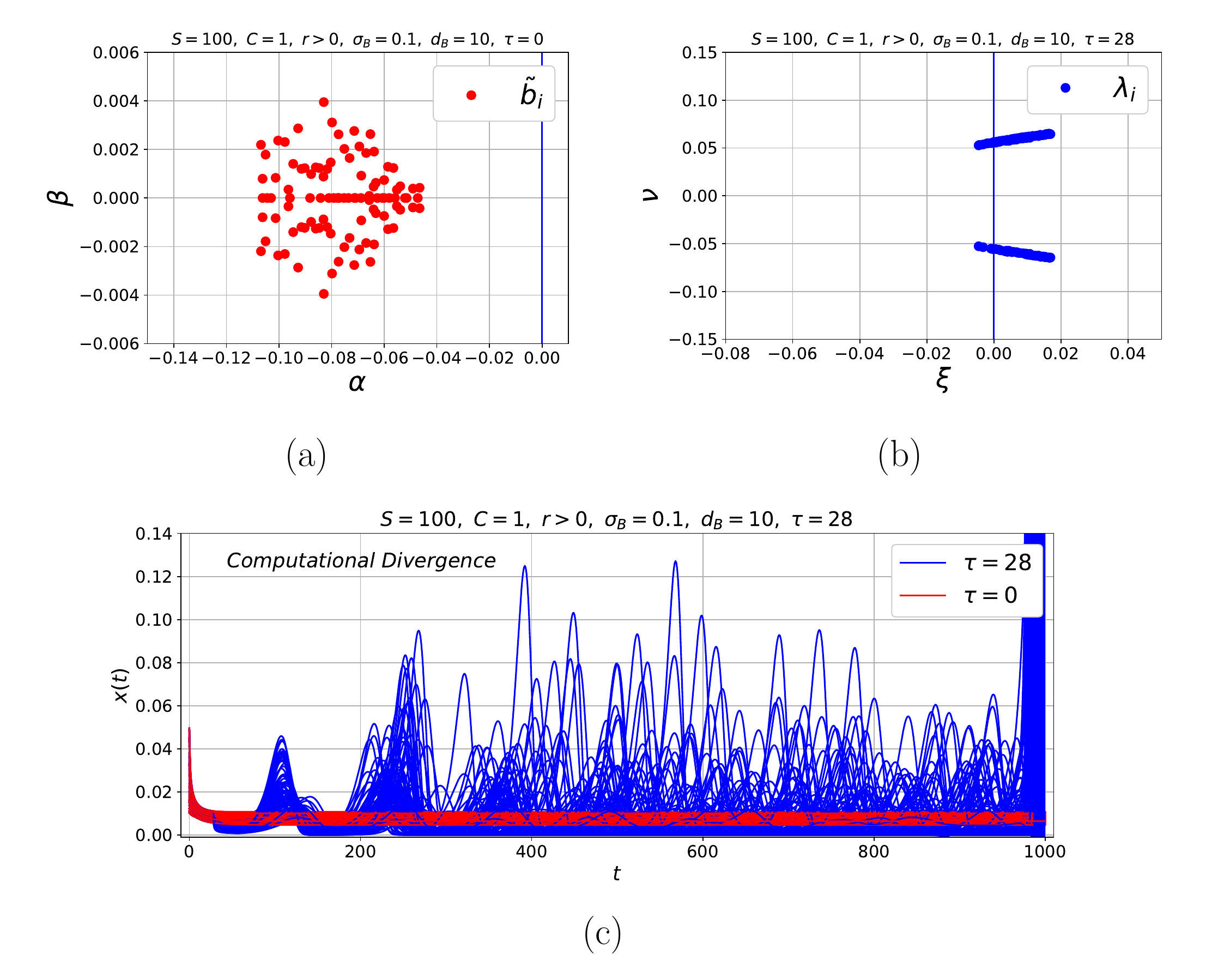}
\caption{
Characterization of a diverging phase for Eq.\eqref{eq:dynamics}. (a) Eigenspectrum of the community matrix $J=\text{diag}(-B^{-1}r)B$ for the non-delayed GLV system. (b) Eigenspectrum of the delayed GLV system for $\tau=28$ (see Eq.~\ref{Lambda_Lambert}). While in panel (a) the real part of the leading eigenvalue is negative, it is instead positive in panel (b). Thus, as depicted by the results of the numerical integration of Eq.\eqref{eq:dynamics} in panel (c), the system without delay is asymptotically stable (red line), whereas it displays a divergence in the presence of delay (blue line). Note, in all panels, $A=0$ and $B$ is a random matrix.
}
\label{nonlin-singleIII}
\end{figure}

The community matrix of Eq.~\ref{CLV} with $\tau=0$ for a feasible equilibrium state is obtained from $J=\text{diag}(-B^{-1}r)B$. The eigenvalues of the community matrix (in the absence of delay) are plotted in the complex plane in panel $(a)$ of Figs.~\mref{nonlin-singleI,nonlin-singleII,nonlin-singleIII}. Being in the coexistence region ($r>0$ ,  $B \in \textbf{DiS}$),  the real parts of the eigenvalues are in the left half-plane (i.e., Re $\lambda<0$). As expected, the trajectories of the system, obtained from the numerical integration of Eqs.~\ref{CLV},  are asymptotically stable as shown by the red lines in panel $(c)$ of Figs.~\mref{nonlin-singleI,nonlin-singleII,nonlin-singleIII}. 

Now, we turn on the delay by gradually by increasing $\tau$. The eigenvalues of the community matrix in the presence of delay are obtained from Eq.~\ref{Lambda_Lambert}, and are plotted in panel $(b)$ of Figs.~\mref{nonlin-singleI,nonlin-singleII,nonlin-singleIII}. Before a certain $\tau_{c}$, the real part of the leading eigenvalue is still in the left half-plane (Re $\lambda<0$), and the trajectories are still asymptotically stable, as shown by the blue lines in panel (c) of Fig.~\mref{nonlin-singleI}. In this regime, the resilience of the system (as measured by the absolute value of the leading eigenvalue of the community matrix) decreases by increasing the delay. As $\tau$ passes through the $\tau_{c}$, the two conjugate leading eigenvalues cross the vertical line given by Re $\lambda=0$, thus confirming a Hopf bifurcation. This is evident by comparing panel (b) of Fig.~\ref{nonlin-singleI} and Fig.~\ref{nonlin-singleII}. This is confirmed by the trajectories of the dynamics which present persistent periodic oscillations as shown by the blue lines in panel $(c)$ of Fig.~\mref{nonlin-singleII}.

Remarkably, $\tau_c$ can be calculated analytically by means of linear stability analysis. Indeed, if we linearize the system around the feasible equilibrium $x^\star=-B^{-1}r$, the system eigenvalues $\lambda_i$ satisfies the following (see Eq.\eqref{Lambda})
\begin{equation}
    \lambda_i = j_i e^{-\tau \lambda_i},
    \label{eq:lambdai_b}
\end{equation} where $j_i$ are the eigenvalues of the community matrix $J$. From this equation it is possible to calculate the critical delay 
directly from the eigenspectrum of the community matrix (see Appendix \textbf{C}) as
\begin{equation}
    \tau_c = \min_{j\in \text{eig}(J)}  \frac{1}{|j|} \arctan{\left|\frac{\text{Re}(j)}{\text{Im}(j)}\right|},
    \label{eq:tauc}
\end{equation}
and the corresponding critical oscillation frequency as
$
\nu_c =\left|\operatorname*{arg\,min}_{j\in \text{eig}(J)}\frac{1}{|j|} \arctan{\left|\frac{\text{Re}(j)}{\text{Im}(j)}\right|}\right|
$. 


By increasing $\tau$ above $\tau_{c}$, the amplitude of the trajectories increase, as can be seen by comparing the blue trajectories in panel $(c)$ of Figs.~\mref{nonlin-singleI,nonlin-singleII,nonlin-singleIII}. When $\tau$ is much larger than $\tau_{c}$ the amplitude of the oscillations becomes so large that it gives rise to numerical divergences which we were not able to fix (see panel $(c)$ of Fig.~\mref{nonlin-singleIII}.). This occurred at some threshold $\tau_c^C > \tau_c$, which did not change a lot by improving the accuracy of the Euler method.

\begin{figure}[ht]
\centering
\includegraphics[width=1\linewidth]{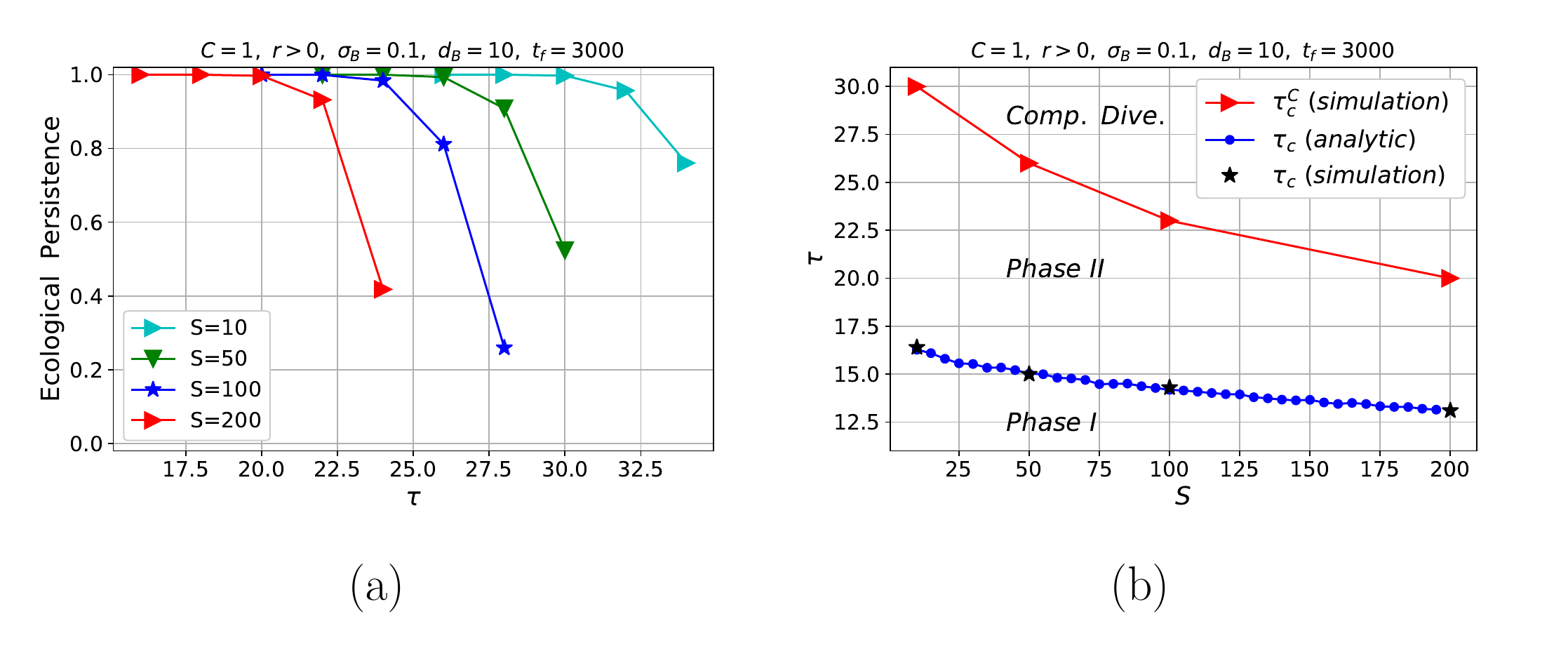}
\caption{
(a) Ecological persistence, i.e., probability of no divergence and no extinction, as a function of delay $\tau$ and for different system size $S$ and (b) stability phase diagram for the system defined by Eq.\eqref{eq:dynamics}. In the latter, Phase I corresponds to the stable phase, phase II to the oscillatory regime, whereas the Computational Divergent phase to the divergent one. The phase separation line $\tau_c$ between the stable and the oscillatory regime is obtained both by means of simulations (black stars, see Eq.\eqref{eq:dynamics}) and analytically (blue line, see Eq.\eqref{eq:tauc}). The second phase separation line (red line) represents instead $\tau_c^C$, which is defined as the maximum $\tau$ below which the system is almost surely ecologically persistent. In both panels, we have considered the intrinsic growth rates $r_i$ to be uniformly distributed in $\mathcal{U}$[0.05,0.1] and the interaction $B$ to be a random matrix, with constant diagonal $d_B=10$ and off-diagonal elements normally distributed with zero mean and $\sigma_B=0.1$. The initial conditions $x_i(0)$ are drawn from an uniform distribution $\mathcal{U}$[0.01,0.05]. Note also, in all panels $A=0$.
}
\label{lin-nonlin-phases}
\end{figure}

We plot in panel (b) of Fig.~\ref{lin-nonlin-phases} the critical time delay for the bifurcation, $\tau_{c}$, and the critical time delay for the divergence, $\tau_c^C$, as a function of diversity, $S$. The curves of $\tau_c$ and $\tau_c^C$ are monotonously decreasing. This identifies two regimes: all the trajectories associated with the region below the blue line are asymptotically stable; the trajectories associated with the region between the blue and the red lines indicate the region of the oscillatory regime (non-point attractors) which depends on $r$ and $B$. The dynamics above the red line leads to (numerical) divergences.


In conclusion, by means of numerical simulations and analytical calculations, we have confirmed the existence of a Hopf bifurcation at $\tau_{c}$ for a delayed GLV with a large number of species. This bifurcation is also observed for the case of the stable but partially feasible equilibrium state (violation of the coexistence condition). However, ecologically speaking, this regime is not interesting to investigate except for the case of invasion of species, which is not the purpose of this study.



\subsection{A more empirical perspective of stability.}
Local stability is an instructive theoretical concept, but very difficult to measure in real ecological scenarios. First of all the underlying assumption is that the ecosystem in its unperturbed state is considered at the equilibrium. However, a real ecosystem is continuously changing, exchanging fluxes of energy with the environment and among species. In other words, a real ecological community is either in stationarity or out of equilibrium system. For these reasons, many field ecologists do not use the stationary-based measures of stability, rather they apply the concept of variability (e.g. the variance of population densities over time, or  the coefficient of variation (CV) of the populations) as indicator of the ecosystem stability \cite{pimm1984complexity,mccann2000diversity}, i.e. the less the variability the more stable the ecological community. 


On this perspective, 
some experimental field works on plant biodiversity \cite{tilman1994biodiversity,tilman1996productivity,tilman1996biodiversity,van1998mycorrhizal} have shown that the diversity within an ecosystem tends to be correlated positively with the community-level stability (measured as the inverse of the CV
in community biomass)  while it is only weakly correlated with  CV for year to year variation in biomass of individual species (e.g. see Figures (6) and (9) in \cite{tilman1996biodiversity}). Also, some other studies on controlled microcosm experiments  \cite{mcgrady1997biodiversity,mcgrady2000biodiversity,morin1995food,naeem1997biodiversity,naeem1998species,yachi1999biodiversity} have suggested that individual specie population-level variation is relatively uninfluenced by diversity,
whereas community-level variations tends to decrease with increased diversity.

\begin{figure}[ht]
\centering
\includegraphics[width=1\linewidth]{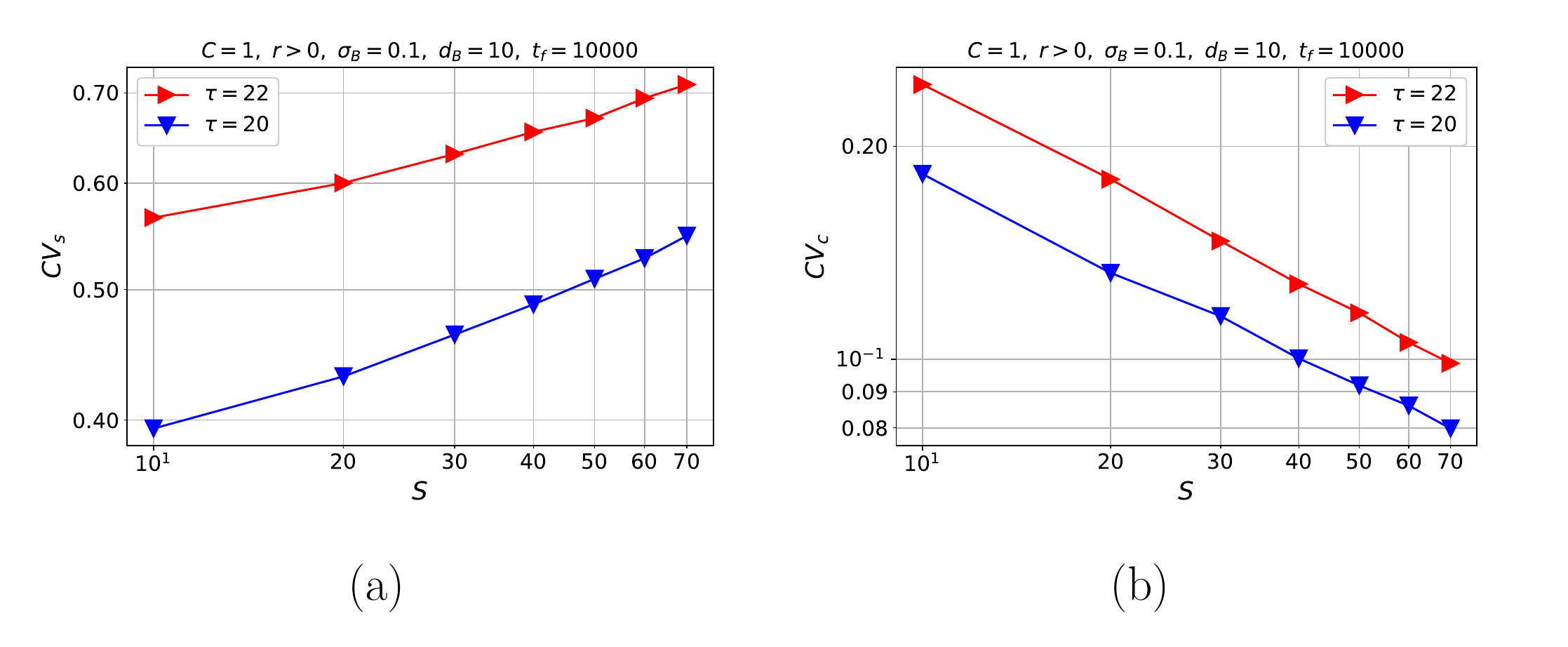}
\caption{Population-level (a) and community-level (b) variability as a function of the system size $S$ for a delayed GLV with $\tau=20$ (blue lines) and $\tau=22$ (red lines). Interestingly, the average coefficient of variation at the individual level $CV_S$ (a) increases with the system size, in agreement with May's result \cite{may1974time}, whereas the coefficient of variation at the community level $CV_c$ reduces as a function of $S$, pointing out that the diversity and the stability are positively correlated, consistently with experimental  \cite{tilman1996biodiversity}.
}
\label{CV}

\end{figure}

We thus want to investigate whatever this relation holds also in our GLV dynamics. Let us highlight that, on this respect, delay is a fundamental, ecological relevant and sufficient mechanism allowing for the emergence of oscillatory regime, at thus of variability at stationarity at both the species and community level. In our context we consider the variations of the populations at the individual species and at the community-level as respectively
\begin{eqnarray}
    CV_s&=&\sum_{i=1}^S\frac{\sqrt{(\sum_{t=T}^{t_f}(x_i(t)-\bar{x}_i)^2)/(t_f-T)}}{S\bar{x}_i}, \cr  CV_c&=&\frac{\sqrt{(\sum_{t=T}^{t_f}(N(t)-\bar{N})^2)/(t_f-T)}}{\bar{N}},
\end{eqnarray}
where $T$ the time step when stationarity is reached, $t_f$ is the length of the simulated time series (here $T=0.1t_f$),
\begin{eqnarray}
    \bar{x}_i&=&\frac{\sum_{t=T}^{t_f}x_i(t)}{t_f-T}, \cr  N(t)&=&\sum_i^Sx_i(t), \cr
    \bar{N}&=&\frac{\sum_{t=T}^{t_f}N(t)}{t_f-T}.
\end{eqnarray}
We note that in the equilibrium regime ($\tau< \tau_c$) $CV_s=CV_c=0$. While, in the oscillatory regime of delayed GLV, we can plot the $CV_{s}$ and $CV_{c}$ as a function of the diversity, respectively. As evident from panels (a) and (b) of Fig.~\ref{CV}, $CV_{c}$ decreases by increasing the diversity, while the $CV_{s}$ increases by increasing the diversity.  In other word, in this out of stationary approach, we find the emergence of a positive diversity-stability relationship as measured by the community level variability,in agreement with experimental observations.


\section{Discussion and Conclusion}

In spite of the relevant theoretical efforts to better understand the relationship between stability and diversity, the celebrated complexity-stability paradox is far from being settled. However, more and more studies highlight many ecological mechanisms that may allow for the emergence of stable ecological communities, where several species coexist. In this work, we have investigated the role of the non-linearity induced by the delay in how species interactions affect species growth rates. We have thus incorporated the delay in a Generalized Lotka Volterra model's ecosystem and considered a null ecological  ecosystem  with  random quenched interactions. 

First, we have found  an analytical connection between the feasibility, stability and diversity of the non-delayed GLV. Then, by gradually increasing the delay, we have numerically observed the emergence of a new dynamical regime. Actually, beyond the asymptotically stable regime where the species reach equilibrium points, we have found an oscillatory regime for delays larger than a critical value. We have also calculated analytically the critical delay which is in very good agreement with numerical simulations. For even larger delays, numerical instabilities lead to the divergence of populations trajectories, although we have some analytical insights suggesting that the true analytical trajectories should be bounded. All in all, our results confirm that delay is detrimental for local GLV stability.

Finally, by employing the variability of oscillations in the population dynamics induced by the delay, we can change perspective and go beyond local stability, to investigate a stability framework more suitable for comparisons with experimental data. Such framework holds also in non-equilibrium regimes by defining stability as the inverse of the coefficient of variation of the ecosystem population dynamics. Consistently with experimental results we find that the variability of the community-level decreases by increasing ecosystem diversity. This suggests new ways to consider the role of the delay in ecological dynamics, moving beyond the equilibrium framework of local stability.
\section{Acknowledgments}
A.M. and E.P. are supported by ``Excellence Project 2018'' funded by Cariparo foundation.  S.A., S.S. and A.M also acknowledge INFN for Lincoln grant and UNIPD DFA BIRD2021 grant. E.P. acknowledges a fellowship funded by the Stazione Zoologica Anton Dohrn (SZN) within the SZN-Open University Ph.D. program. M.S. acknowledges  the fellowship from the Departemant of Physics And Astronomy ‘G. Galilei’, University of Padova. Also, M.S. acknowledges Dr. M. Sadeghi and the support of the Iran National Science Foundation (INSF) (grant number: 98023811)

\begin{center}
\textbf{\large Appendix A}
\end{center} \ \\
\setcounter{equation}{0}
\setcounter{figure}{0}
\setcounter{table}{0}
\makeatletter
\renewcommand{\theequation}{A\arabic{equation}}
\renewcommand{\thefigure}{A\arabic{figure}}
Here we provide some sufficient conditions for stability of Eq.~\ref{CLV}, i.e., for the classical GLV model without delay. 
\subsection{Linear Stability Analysis}
Let $H=\{1,2,...,S\}$ be the set of the species present in the system. If we consider a partially feasible equilibrium $x^\star$, i.e., an equilibrium that could have a positive number of extincted species, we can denote $P$ and $Q$ as two subsets of $H$ such that $i\in P$ if $x_i^\star>0$ and $i\in Q$ if $x_i^\star=0$. Naturally, when $H=P$ the equilibrium is feasible. Let us also define $x_i=u_i+x_i^\star$, where $u$ is a small perturbation from the equilibrium state $x^\star$. Then, the linearized form of the Eq.~\ref{CLV} reads 
\begin{eqnarray}
\dot{u}&=&Mu
 \label{LinCLV} 
\end{eqnarray}
where the rearranged community (Jacobian) matrix yields
\begin{eqnarray}
M=
\left(\begin{array}{c|c}
   x^\star_i\frac{\partial g_i}{\partial x_j } & x^\star_i\frac{\partial g_i}{\partial x_k } \\
   \hline
   0 & \delta_{hk}g_k \\
\end{array}\right)_{x^\star},
\label{M_LS}
\end{eqnarray}
where $i,j \in P$ and $h,k \in Q$ and $\delta_{hk}$ is the Kronecker delta function, which is $1$ if $h=k$, and $0$ otherwise. The subscript of the round bracket means that all the partial derivative are computed in $x^\star$. Let us also call the upper left block of $M$ as 
\begin{eqnarray}
K=x_i^\star\Bigg(\frac{\partial g_i}{\partial x_j }\Bigg)_{x^\star}.
 \label{ESg} 
\end{eqnarray}

\textbf{Theorem 1} \cite{goh1978sector}: The partially feasible equilibrium $x^\star$ of  Eq.~\ref{CLV} is locally stable if all the eigenvalue of the $K$ have negative real part and
\begin{eqnarray}
g_k(x^\star)<0, \quad \text{for all } \quad k\in Q.
\end{eqnarray}
We denote here by $\textbf{LS}$ the set of linear stable matrices. As apparent in Fig.~\ref{containment_relation}, this condition is the superset of all the forthcoming sufficient conditions for the stability of Eq.~\ref{CLV}. Note, this condition works for any general form of Eq.~\ref{eq:LV1}. 
\subsection{D-stability: a guarantee for local stablity}
In the case of Lotka-Volterra model Eq.~\ref{CLV}, if a feasible equilibrium state exists and $A$ is invertible, using Eq.~\ref{M_LS}, the community matrix reads
\begin{eqnarray}
M=\text{diag}(x^\star)A,
 \label{Mfea} 
\end{eqnarray}
where
\begin{eqnarray}
x^\star=-A^{-1}r.
 \label{xfea} 
\end{eqnarray}
In this case, the stability of the community matrix $M$ is obtained from \textbf{Theorem 1}. Since $M$ is simply the multiplication of the interaction matrix $A$ times the positive diag($x^\star)$, which is in turn a function of (the inverse of) $A$ and the intrinsic growth rates $r$. This then rises the question about what conditions on $A$ do grant the stability of $M$ for any positive $\text{diag}(x^\star)$. This is the reason behind the D-stability condition. 

\textbf{Definition 1} \cite{arrow1958note}: A matrix $M\in R^{S\times S}$ is said to be $D$-stable, if $DM$ is stable ($DM \in \textbf{LS}$), where $D$ is a diagonal matrix with positive diagonal elements $d_{ii} > 0$.

We denote here by $\textbf{DS}$ the set of D-stable matrices. In the containment relation we can assert $\textbf{DS} \subset \textbf{LS}$. As noticed, characterizing the $D$-stability is not trivial. However, as a testable necessary condition for the definition of $D$-stability, we bring here \textbf{Theorem 2} as

\textbf{Theorem 2} \cite{quirk1965qualitative,logofet2005stronger}: Any $D$-stable matrix is a $-P^{0}$, or, in formal terms, $\textbf{DS} \subset -P^{0}$. 
Where

\textbf{Definition 2} \cite{johnson1974sufficient}: A matrix $M\in \textbf{R}^{S\times S}$ is said to be $P^{0}$-matrix if all principal minors of $M$ are non-negative and if for each order $k = 1,..., S$, at least one $k$ by $k$ principal minor is positive.

\textbf{Definition 3} \cite{takeuchi1980existence}: A matrix $M\in R^{S\times S}$ is said to be $P$-matrix if all principal minors of $M$ are positive.

To check the $D$-stability of the system of $S \leqslant 4$, there exist some testable conditions in \cite{kanovei1998d,logofet2005stronger}.

\subsection{Total stability: a guarantee for species deletion stability}
Disappearing (or deleting) the species that take place not because of the population dynamics can be considered as a relatively large perturbation on population dynamics \cite{pimm1982food}. If $M$ is interpreted as a community matrix, one can argue if there exist some conditions on the $M$ that do grant the stability of the community under the perturbation of species deletion. Here, after defining the concept of ``species deletion stability'', we present a sufficient condition to grant this stability and then we present a necessary condition for this sufficient condition. 

\textbf{Definition 4} \cite{pimm1982food}: A system is said to be species deletion stable if, following the removal of a species from the system, all of the remaining species are retained at a new, locally stable equilibrium.

One sufficient condition to grant the ``species deletion stability'' is the condition of ``Total stability'' 

\textbf{Definition 5} \cite{quirk1965qualitative}: A matrix $M$ is said to be totally stable if every principal subset of $M$ (i.e., every sub-set whose determinant is a principal minor of $M$) is D-stable.

We denote here by $\textbf{TS}$ the set of Total stable matrices. The Definition 5 implies that the Total stability is a subset of $D$-stability, i.e.,  $\textbf{TS}\subset\textbf{DS}\subset \textbf{LS}\subset \textbf{S}$. One necessary condition to grant Total stability is

\textbf{Theorem 3} \cite{logofet2005stronger}: Any Total stable matrix is a $-P$, or, in formal terms, $\textbf{TS} \subset -P$.

If $M$ is interpreted as the interaction matrix, one very important consequence is that if $M \in -P$, for every $r \in \textbf{R}^S$, there exist a unique  non-negative equilibrium state for Eq.~\ref{CLV} \cite{murty1972number,takeuchi1980existence}. However, the stability of that is not granted. In theorem \textbf{Theorem 4}, the condition for stability of that will be presented.

\subsection{Lyapunov diagonal stability; a guarantee for global asymptotic stability }
The existence of an unique domain (or basin) of attraction for an equilibrium state of Eq.~\ref{CLV} grants the global asymptotic stability of the trajectories of the populations. At this stage, we can look at the conditions on the interaction matrix $A$ that grant the global stability of the ecological community. To this purpose, we need to define the concept of Lyapunov diagonal stability. We denote here by $\textbf{DiS}$ the set of Lyapunov diagonal stable matrices. 

\textbf{Definition 6} \cite{kraaijevanger1991characterization} . When $M$ is an $n \times n$ real matrix, $M \in \textbf{DiS}$ implies that there exists an $n \times n$ positive definite diagonal matrix $D$ such that $DM + M^{\dagger}D$ is negative definite.

\textbf{Theorem 4} \cite{takeuchi1980existence}: If $M \in \textbf{DiS}$, then the system defined by Eq.~\ref{CLV} has a non-negative and stable equilibrium point for every intrinsic growth rate $r \in \textbf{R}^S$. 

This important theorem is a direct consequence of Lyapunov functions \cite{goh1978sector,takeuchi1980existence} and linear complementary theory \cite{murty1972number} that have been successfully applied to show global stability of Lotka-Volterra systems.

There is a great deal of importance considering the existence of multiple domains of attractions for ecological systems \cite{gilpin1976multiple}. Theorem 4 gives a class of systems that do not have multiple domains of attractions. This class \textbf{DiS} is defined only in terms of the interaction matrix $A$ and does not involve the intrinsic growth rate $r \in \textbf{R}^S$. 

Consequently, if the interaction matrix $A$ of a community being negative definite, every feasible equilibrium points are globally stable. Also, any principle matrix of a negative definite matrix is negative definite. Thus, any feasible point of a reduced system is globally stable. As final statement here, the containment relation goes like  $\textbf{DiS}\subset\textbf{TS}\subset\textbf{DS}\subset \textbf{LS} \subset \textbf{S}$. There are some other testable conditions that can be found in the beautiful and inspiring work of O. Logofet \cite{logofet2005stronger}, such as Quasi-Dominant stability and Qualitative Stability \cite{may1973qualitative}.  

\begin{figure}[ht]
\centering
\includegraphics[width=.6\linewidth]{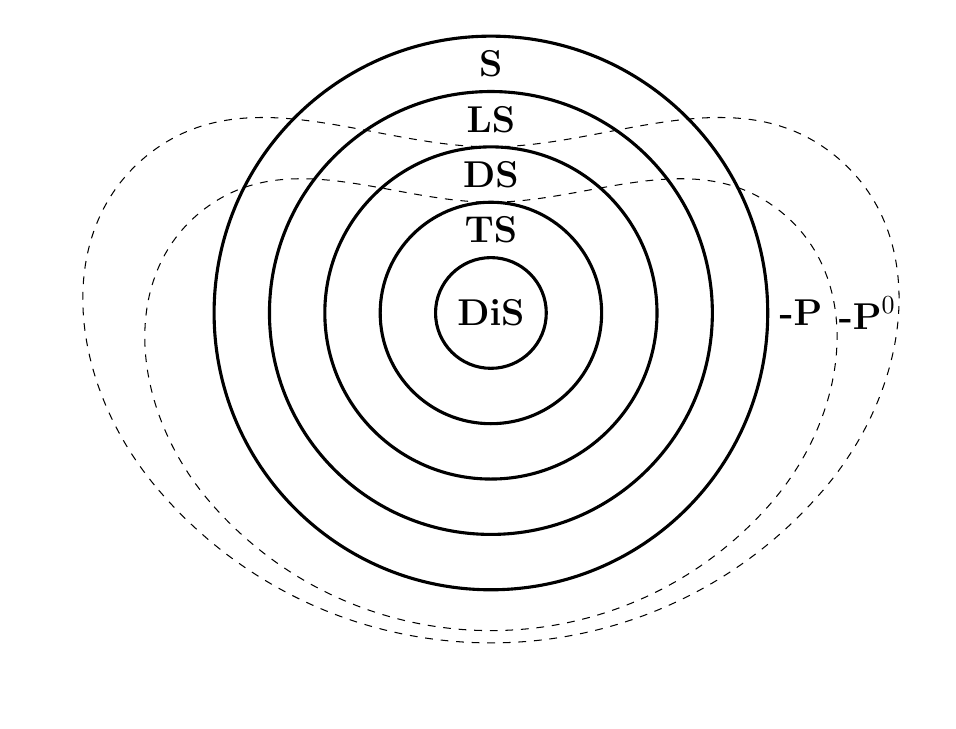}
\caption{Schematic containment picture of stability conditions. The \textbf{S} stand for stability. The sufficient conditions, in order of inclusion, are Linear stability (\textbf{LS}), D-stability (\textbf{DS}), Total stability (\textbf{TS}), and Diagonal Lyaponov stability \textbf{DiS}. The necessary conditions are $P-$matrix and $P^0$-matrix. The curves of sufficient (necessary) conditions are drawn by solid (dashed) lines. The outer (inner) dash curve indicate a necessary condition for D-stability (Total stability).}
\label{containment_relation}
\end{figure}

To sum up this section, we plot the schematic containment picture of  all aforementioned stability conditions in Fig.~\ref{containment_relation}.

\ \\
\quad 
\begin{center}
\textbf{\large Appendix \textbf{B}}
\end{center}
\setcounter{equation}{0}
\setcounter{figure}{0}
\setcounter{table}{0}
\makeatletter
\renewcommand{\theequation}{B\arabic{equation}}
\renewcommand{\thefigure}{B\arabic{figure}}

\section{Stability and feasibility for the GLV without delay}
As already stated in the main text, it is possible to have an insight on the feasibility and the stability of the system's equilibria for the GLV model without delay, at least in some special cases. It is worth highlighting that quantifying the feasibility properties is a fundamental step in the study of such systems. Indeed, we remark that studying the non trivial equilibrium $x^\star=-A^{-1}r$ is meaningful only if it is feasible. On the other hand, when at least one component $x_i^\star$ is negative, $x^\star$ does not give any clue neither to which species would become extincted, nor to the number of surviving species. Let us clarify this concept by mean of a simple but paradigmatic example with $S=2$ species. For instance, let the system of ODEs be
\begin{equation}
\begin{cases}
 \dot{x}_1 &= x_1 \left(1-x_1-2 x_2\right) \\
 \dot{x}_2 &= x_2 \left(1+ \frac{3}{4}x_1 - x_2\right)
\end{cases},
\end{equation} whose equilibria are $(0,0)$, $(0,1)$, $(1,0)$, and the non-trivial, unfeasible $x^\star=\left(-\frac{4}{10},\frac{7}{10}\right)$. Although one can naively think that the system would evolve towards $(0,1)$ since $x_1^\star<0$ and $x_2^\star>0$, rather interestingly, $(1,0)$ turns out to be the stable, attracting equilibrium. In the same way, it is not hard to construct a 3-species model where $x^\star$ has two negative components, but the stable equilibrium can have two surviving species.

\subsection{Feasibility}
Although a general characterization of the feasibility for the GLV model, it is nevertheless possible to quantify the feasibility probability when $A$ is a diagonal-dominant random matrix, i.e. when $d_A\gg\sigma_A$. In particular, if we define $M$ as the off-diagonal part of $A$, i.e., as $A = -d\, \mathds{I} + M$, then we can formally expand the inverse of the interaction matrix as follows:
\begin{equation}
A^{-1} = -\frac{1}{d_A} \sum_{k=0}^{\infty}\left(\frac{M}{d_A}\right)^k = -\frac{1}{d_A}  \left(\mathds{I} + \frac{M}{d_A}\right) + o\left(\frac{\sigma_A}{d_A}\right).
\end{equation}
Under these hypotheses, at the first order, the steady states now reads
\begin{equation}
x^\star_i = \frac{r_i}{d_A} + \sum_{j=1}^S \frac{M_{ij}r_j}{d_A^2} + o\left(\frac{\sigma_A}{d_A}\right),
\label{eq:xstarapp}
\end{equation}
while the feasibility condition naturally becomes
\begin{equation}
r_i + \sum_{j=1}^S \frac{M_{ij}r_j}{d_A} > 0 \quad \forall \, i=1,\dots,S.
\label{eq:feasibility0}
\end{equation}
We remark the importance of Eq.~\ref{eq:xstarapp}: when $d_A$ is large, if $A$ and $r$ are random variables as defined in the main text, the steady states populations are distributed around the mean value
\begin{equation}
E[x^\star_i] = \frac{E[r]}{d_A} 
\label{eq:ex}
\end{equation}
with variance
\begin{equation}
Var[x^\star_i] = \frac{1}{d_A^2}\left(Var[r] + \frac{C(S-1)\sigma^2}{d_A^2}\left(E[r]^2+Var[r]\right)\right).
\label{eq:varx}
\end{equation}

This result is general and does not depend on the particular choice of the distribution for $A$ and $r$. If we now additionally assume that the growth rates are all equal and in particular--since a rescaling of the modulus of $r$ does not affect the feasibility--that they are equal to 1, then the steady state components are randomly distributed with mean and variance given by
\begin{equation}
\begin{aligned}
E[x_i] &= \frac{1}{d_A}, \\
Var[x_i] &= \frac{C(S-1)\sigma^2}{d_A^4}.
\end{aligned}
\end{equation} This shows that the mean value of $x^\star_i$ is always positive and depends only on $d_A$, the other parameters $S$, $C$ and $\sigma$ being involved only in the spreading around it. In particular, the larger the system size $S$ or the connectivity $C$ of the interactions, the more spread are the $x^\star_i$, the more likely some components are negative. Indeed, as reported in the main text, a good estimator for the probability for the feasibility condition to hold is given by the quantity
\begin{equation}
\int_0^{+\infty} \mathcal{N}\left(x_i\,\lvert\, E[x_i] ,\, Var[x_i]\right) =  \frac{1}{2}\text{erfc}\left[-\frac{d_A}{\sqrt{2C\left(S-1\right)}\sigma}\right].
\label{eq:feas0-app}
\end{equation}

 This can be seen also considering explicitly the feasibility condition, given by Eq.~\ref{eq:feasibility0}, which in this case becomes
\begin{equation}
1 + \sum_{j=1}^S \frac{M_{ij}}{d_A} > 0 \quad \forall \, i=1,\dots,S.
\end{equation}
Defining the random variable $z_i$ as
$z_i \equiv \sum_{j=1}^S M_{ij}$, we are interested in estimating its probability density function $p\left(z_i\right)$, so that the feasibility condition translates in computing the probability
\begin{equation}
\text{Prob}\left(z_i > -d \right) = \int_{-d}^{+\infty} p(z_i)\,dz_i.
\end{equation}
This can be easily computed by mean of the law of total probability
\begin{equation}
p(z_i) = \sum_{N=0}^{S-1} p(z_i\rvert N)\, p(N);
\end{equation}
where $p(N)$--the probability that the $i$-th row of $M$ has exactly $N$ non-zero distributed entries over the $S-1$ off-diagonal elements--is a binomial distribution, since the outcome in each entry of $M$ is a Bernoulli trial with probability $C$:
\begin{equation}
p(N) = C^N \left(1-C\right)^{S-1-N} \binom{S-1}{N};
\end{equation}
while 
\begin{equation}
p(z_i\rvert N) = \begin{cases}
\delta(z_i) &\text{if } N=0 \\
\mathcal{N}(z_i\,\lvert\,0,\,N\sigma_A^{2}) &\text{if } N > 0,
\end{cases}
\end{equation}
where the factor $N$ in the variance comes from the fact that $z_i$ is the sum of $N$ normally distributed terms with variance $\sigma^2$.

Therefore,
\begin{equation}
\begin{split}
\text{Prob}\left(z_i > -d \right) &=  \int_{-d}^{+\infty} dz_i \sum_{N=0}^{S-1}  p(N) p(z_i\rvert N)\, \\
&= \left(1-C\right)^{S-1} + \sum_{N=1}^{S-1} C^N \left(1-C\right)^{S-1-N} \binom{S-1}{N}  \int_{-d}^{+\infty} dz_i \,\mathcal{N}(z_i\,\lvert\,0,\,N\sigma^{2}) \\
&= \left(1-C\right)^{S-1} + \sum_{N=1}^{S-1} C^N \left(1-C\right)^{S-1-N} \binom{S-1}{N} \frac{1}{2}\text{erfc}\left[-\frac{d}{\sqrt{2N}\sigma_A}\right]
 \end{split}
 \label{eq:feasibility1}
\end{equation}
and since this condition should hold for each row $i$,
\begin{equation}
\text{Prob}\left(\text{feasibility} \rvert S,d,c,\sigma_A\right) = \left[\text{Prob}\left(z_i > -d_A \right)\right]^S.
\end{equation}

The main advantage of this equation is its closed-form fashion. On the other hand, it becomes computationally hard to be computed for large $S$. Furthermore, the dependence of the feasibility probability on the model parameters is not immediately clear within this formulation. For this purpose, two approximations can be performed, thanks to the Central Limit Theorem.

The first approximation--which holds in the limit of large $S$--is simply to approximate the binomial distribution inside Eq.~\ref{eq:feasibility1} with a normal one, with mean $\mu_N= C\left(S-1\right)$ and variance $\sigma_N^2=\left(1-C\right)C\left(S-1\right)$. 

The second one, instead, is valid in the limit of large number of non-zero entries of $M$, i.e., in the limit of large $cS$. In this limit, we can approximate the probability density function $p\left(z_i\right)$ with a normal distribution with mean $E[z]=0$ and variance $E[z^2]=\sigma_A^2C\left(S-1\right)$. Indeed,
\begin{equation}
E[z] =  \int_{-\infty}^{+\infty} dz\, p(z) \, z= \sum_{N=0}^{S-1}p(N)   \int_{-\infty}^{+\infty} dz\, \underbrace{p(z\rvert N)}_{\text{even}}\, \overbrace{z}^{\text{odd}} = 0
 \end{equation}
and
\begin{equation}
\begin{split}
E[z^2] &=  \int_{-\infty}^{+\infty} dz\, p(z) \, z^2 =  p(0)\int_{-\infty}^{+\infty} dz\, \delta\left(z\right) \, z^2 + \sum_{N=1}^{S-1}p(N)  \int_{-\infty}^{+\infty} dz\, ,\mathcal{N}(z\,\lvert\,0,\,N\sigma_A^{2})  \\
&= 0 +  \sum_{N=1}^{S-1}p(N) N \sigma_A^2 = \underbrace{C\left(S-1\right)}_{\mu_N} \sigma_A^2. \\
\end{split}
\end{equation}

Thus, we recover exactly the quantity shown in Eq.~\ref{eq:feas0-app}, that is,
\begin{equation}
\text{Prob}\left(z_i > -d_A \right) =   \int_{-d}^{+\infty} dz_i\, \mathcal{N}(z_i\,\lvert\,0,\,C\left(S-1\right)\sigma_A^{2}) = \frac{1}{2}\text{erfc}\left[-\frac{d_A}{\sqrt{2C\left(S-1\right)}\sigma_A}\right]
\label{eq:feasibility2}
\end{equation}
and since this condition should hold for each row $i$, again the feasibility probability can be obtained by raising this result to the power of $S$. 

\subsection{Stability}
As already stated in the main text, the local stability of an equilibrium can be studied by mean of the eigenspectrum of Jacobian evaluated around it. To this purpose, let us consider the rather general system of equation
\begin{equation}
    \dot{x}_i(t)=x_i(t)\,g_i\left[x(t)\right] \equiv f_i\left[x(t)\right],
\end{equation} where $g_i$ is in general a non-linear function of the abundances $x_i(t)$. If we focus on a non-trivial equilibrium $x^\star$, i.e. on a equilibrium satisfying the condition $g\left[x^\star\right]=0$, then its Jacobian (also known as interaction matrix) reads
\begin{equation}
    J_{ij}=x_i^\star \left.\frac{\partial g_i}{\partial x_j}\right|_{x^\star}.
\end{equation}
For instance, for the classical GLV model, where $g\left[x\right]=r+Ax$, the community matrix thus simply reads $J_{ij}=x_i^\star a_{ij}$.
If we hypothesise that this matrix has a diagonal, dominant part, and an off-diagonal, subdominant part, i.e. supposing that $\left|\left.\frac{\partial g_i}{\partial x_i}\right|_{x^\star}\right| \gg \left|\left.\frac{\partial g_i}{\partial x_j}\right|_{x^\star}\right| \, \forall \, j\neq i$, then we can decompose the interaction matrix as
\begin{equation}
J_{ij} = \delta_{ij} x^\star_i  \left.\frac{\partial g_i}{\partial x_j}\right|_{x^\star} + \left(1-\delta_{ij}\right)x^\star_i  \left.\frac{\partial g_i}{\partial x_j}\right|_{x^\star}  \equiv J^0_{ij} + V_{ij},
\end{equation}
where the eigenvalues of the diagonal matrix $J^0$ are obviously $\lambda_i^0=J^0_{ii}=x^\star_i  \left.\frac{\partial g_i}{\partial x_j}\right|_{x^\star}$ and its eigenvectors $u^{(i)}$ have components $u^{(i)}=\delta_{ij}$. If $\lambda_i$ is the $i$-th eigenvalue of $J$, we can thus exploit perturbative expansion techniques for estimating the eigenvalues of the interaction matrix:
\begin{equation}
\begin{split}
\lambda_i &= M^0_i +  u^{(i)}\cdot V u^{(i)} + \sum_{j\neq i} \frac{\left(u^{(i)}\cdot V u^{(j)}\right)\left(u^{(j)}\cdot V u^{(i)}\right)}{M^0_i-M^0_j} + \text{h.o.t.} \\
&=x_i^\star \left.\frac{\partial g_i}{\partial x_i}\right|_{x^\star} + \sum_{j\neq i} \left.\frac{\partial g_i}{\partial x_j}\right|_{x^\star} \left.\frac{\partial g_j}{\partial x_i}\right|_{x^\star} \frac{x_i^\star x_j^\star}{x_i^\star - x_j^\star}+ \text{h.o.t.}.
\end{split}
\end{equation}
and the second term is of the second order. Therefore, for the GLV we recover Eq.\eqref{eq:lambdaGLV1} of the main text.





\ \\
\quad 
\begin{center}
\textbf{\large Appendix \textbf{C}}
\end{center}
\setcounter{equation}{0}
\setcounter{figure}{0}
\setcounter{table}{0}
\makeatletter
\renewcommand{\theequation}{C\arabic{equation}}
\renewcommand{\thefigure}{C\arabic{figure}}

\section{Critical delay for a pure delayed GLV}
Let us consider a pure delayed GLV system 
\begin{equation}
    \dot{x}_i(t)=x_i(t)\big(r_i+\sum_{j}b_{ij}x_j(t-\tau)\big).
\end{equation} Thus, the non-trivial equilibrium is $x^\star=-B^{-1}r$ and the linearization around it reads $
    \dot{u}(t)= \tilde{B}\, u(t-\tau)$, where $u(t)\equiv x(t)-x^\star$ and $\tilde{B}=\text{diag}(x^\star )B$ is the community matrix. Plugging the ansatz solution $u_i(t)\propto e^{\lambda_i t}$we get
\begin{equation}
    \lambda_i = \tilde{b}_i e^{-\lambda_i \tau },
    \label{eq:lambda_delay}
\end{equation}
where $\lambda_i$ is the system eigenvalue and $\tilde{b}_i$ is the eigenvalue of the community matrix. By taking the real and the imaginary part of Eq.\eqref{eq:lambda_delay} we obtain the system
\begin{equation}
    \begin{cases}
     \xi &= e^{-\tau \xi} \left(\alpha \cos{\tau \nu} + \beta \sin{\tau \nu}\right) \\
     \nu &= e^{-\tau \xi} \left(\beta \cos{\tau \nu} - \alpha \sin{\tau \nu}\right) \\
    \end{cases},
\end{equation} where we have denoted $\xi=\text{Re}(\lambda)$, $\nu=\text{Im}(\lambda)$, $\alpha=\text{Re}(\tilde{b})$, and $\beta=\text{Im}(\tilde{b})$.
At criticality, $\xi=0$ and we obtain after some manipulations 
\begin{equation}
    \begin{cases}
    \nu^2 = \alpha^2+\beta^2 \\
    \tau_c(\alpha,\beta) = \frac{1}{|\nu|} \arctan{\left|\frac{\alpha}{\beta}\right|}
    \end{cases}.
\end{equation}
Hence, the critical delay $\tau_c$ would simply be the minimum $\tau_c(\alpha,\beta)$ over the eigenspectrum of the community matrix.

\bibliographystyle{unsrt}
\bibliography{Draft}

\end{document}